# Tailoring the band structure of plexcitonic crystals by strong coupling

Fatemeh Davoodi[1, †], Masoud Taleb[1, †], Florian K. Diekmann[1], Toon Coenen[2], Kai Rossnagel[1,3,4], and Nahid Talebi[1,4]*

[1]*Institute of Experimental and Applied Physics, Kiel University, 24098 Kiel, Germany*

[2]*Delmic B.V., Kanaalweg 4, 2628 EB, Delft, The Netherlands*

[3]*Ruprecht Haensel Laboratory, Deutsches Elektronen-Synchrotron DESY, 22607 Hamburg, Germany*

[4]*Kiel, Nano, Surface, and Interface Science, Kiel University, 24098 Kiel, Germany*

†*These Authors contributed equally.*

E-Mail: talebi@physik.uni-kiel.de

**Abstract**

Transition-metal dichalcogenides with their exciton-dominated optical behavior emerge as promising materials for realizing strong light-matter interactions in the visible range and at ambient conditions. When these materials are combined with metals, the energy confining ability of plasmon polaritons in metals below the diffraction limit, allows for further enhancing and tailoring the light-matter interaction, due to the formation of plexcitons in hybrid metal-TMDC structures at the interface. Herein, we demonstrate that the coupling between quasi-propagating plasmons in plasmonic crystals and excitons in $WSe_2$, provides a multi-oscillator playground for tailoring the band structure of plasmonic crystal structures and results in emerging flat bands. The cathodoluminescence spectroscopy and angle-resolved measurements combined with the numerically calculated photonic band structure confirm a strong exciton-plasmon coupling, leading to significant changes in the band diagram of the hybrid lattice and the ability to tailor the band diagram via strong coupling. The hybrid plexcitonic crystal structures investigated here sustain optical waves with remarkably low group velocities. These results could be used for designing tunable slow-light structures based on the strong-coupling effect and pave the way toward plexcitonic topological photonic structures.

1. Introduction

Strong light–matter interaction is a key aspect of future quantum optical systems.[1-2] Controlling the coupling efficiency of quantum emitters to optical cavities continues to be a subject of increasing interest for the observation of various phenomena such as Bose–Einstein condensation[3-4] and nonclassical light generation at room temperatures.[5-6] Plasmonic nanoparticles are open-system nanoresonators,[7-8] that allow for both localization and an efficient coupling to the far-field due to the radiation damping. Confinement of electromagnetic fields to regions well below the diffraction limit,[9] associated with plasmonic systems opens a wide range of applications based on extreme light concentration, including nanophotonic lasers,[10] biochemical sensing,[11] and controlling light at the nanoscale.[12] There has been a remarkable progress and growing excitement about exploring plasmonic devices at quantum level.[13,14-15]

Particularly, two-dimensional plasmonic crystals which are formed by periodically patterned metallic structures (i.e. particle or hole arrays),[16,17] could support hybrid photonic–plasmonic nanocavities and possess reasonably high quality factors and longer lifetimes compared to localized surface plasmons[18] such as disks,[19] spheres,[20] and dimers.[21,22] Plasmonic crystals have been recently used to investigate and exploit the strong light–matter coupling.[23,24] The optical modes of plasmonic crystals; i.e. plasmonic Bloch modes (PBM), have mixed localized and propagation features, that allow for the emergence of polaritons due to their localization on lattice sites, and at the same time transferring the electromagnetic energy by hopping between the lattice sites.

Excitons in quantum dots or semiconductors are prototypes of quasi two-level quantum systems.[25,23] Due to the large binding energy of excitons in semiconducting transition-metal dichalcogenides (TMDCs) at room temperature,[26,27] TMDCs offer an intriguing platform for strong light-matter interactions. Recently it was reported that a thin film of $WSe_2$ with its exciton-dominated optical behavior is a promising material for realizing exciton-polaritons.[28,29] Furthermore, the effective interaction of plasmons and excitons can lead to hybrid quasi-particles called plexcitons.[30,31] The integration of TMDCs with plasmonic crystals may offer a number of applications via tailoring of the plexcitonic band structure which defines emerging dynamics, slow group velocities, and large effective plexciton masses.[32,33] These aspects provide further control knobs for tailoring the dynamics in quasi two-level systems towards single-photon nonlinearities,[34] since they provide a platform for ultraslow and enhanced light-matter interactions. Single-photon nonlinearities are promising for devising active nanophotonic systems with lowest level energy consumptions.[35]

Here, by means of cathodoluminescence (CL) spectroscopy,[36,37] we investigate the spatial and spectral properties of quasi-propagating plexcitons in $WSe_2$/Au-lattice hybrid structures. Using an electron-beam probe, we demonstrate the strong coupling of excitons in thin $WSe_2$ flakes with quasi-propagating PBMs, and reveal the spatial configuration of the emerging plexcitons via scanning the system with electron beams. The plasmonic lattice introduced here is composed of a square lattice of holes incorporated into a thin gold film. Although the interaction between excitons in TMDCs and plasmons in plasmonic lattices has been already researched,[38-39] the effect of the strong coupling on the band structure of the hybrid lattice and the formation of plexcitonic flat bands has been not yet reported. Here, thin semiconducting WSe$_2$ flakes are used owing to their long – lasting excitons, resulting in strong plasmon-exciton couplings as manifested in a Rabi energy splitting of 0.515 eV and the formation of plexcitons whose spatial coherence and propagation mechanisms are investigated with high spatial resolution. Through the numerical and experimental exploration of the band structures, we unravel the emergence of plexcitonic flat bands and demonstrate that the dispersion can be further controlled via the thickness of the WSe$_2$ flakes. Our findings pave the way toward the design and the control of optical functionalities through the strong-coupling of quantum-mechanical oscillators.

2. Results

*Plexciton Formations and flat bands* – The plasmonic crystal considered here presents a square array of circular holes with a diameter of 0.8 µm and a lattice constant of 1.6 µm perforated in a thin gold film with a thickness of 50 nm designed to exhibit resonances at the visible range (Fig. 1a). The relatively large size of the holes, allows for hybrid photonic-plasmonic resonances, since the holes can themselves act as a cavity and host photonic modes confined to the geometrical region within the hole, with the plasmonic region acting as bridge to connect the photonic modes via coupling to the surface plasmons. A square lattice was chosen as it represents one of the more basic implementations of a two-dimensional periodic crystal reducing the complexity in the interpretation of the band structure of emerging plexcitons. Thin WSe$_2$ flakes with various thicknesses are positioned on top of the plasmonic crystal, using liquid exfoliation (See Supporting

Information for more details) (Fig. 1a and b). A scanning electron microscope (SEM) is used to raster scan and probe the surface of the sample with a high-energy electron beam of 30 keV. Inside the SEM, a paraboloid mirror is positioned above the sample to collect and direct the generated CL radiation onto a CCD camera (Fig. 1a). The detector is capable of resolving the collected CL emission as hyperspectral or angle-resolved images.

Due to the spin-orbit interactions in semiconducting $WSe_2$, two energetically distinct types of excitons are excited in this material, namely A and B excitons, with binding energies of 1.68 eV and 2.05 eV, respectively. PBMs can strongly interact with both A and B excitons. Normally, the strong-coupling effect results in the formation of band gaps, due to the mode repulsion between the oscillators, i.e., plasmons and excitons.[40] Here, the interaction between PBMs and both excitons lead to the hybridized PBM-exciton modes and the creation of plexcitons in narrow energy ranges, whereas two distinct bandgaps are formed at the energies of 1.57 eV to 1.63 eV, and 1.84 eV to 2.0 eV, for a $WSe_2$ flake with the thickness of 60 nm positioned on top of the plasmonic crystal (Fig. 1c), as revealed by our numerical calculations. In order to gain deeper insight into the behavior of the hybrid plexcitonic system in comparison with the plasmonic crystal and compare it with the experimental results – that will be presented later, optical modes of the plexcitonic crystal and their dispersion relations are numerically explored at the next step.

For calculating the band diagram, only the first-order Brillouin zone is considered, and the Γ-X-M-Γ path is adopted (see Supporting Information for more details about the numerical schemes employed here). The calculated dispersion diagram of the bare plasmonic crystal (i.e., without $WSe_2$), demonstrates in overall no specific bandgap in the energy range of 1.3 eV-2.2 eV (Fig. 2a). For some points of Brillouin zone, i.e., at $\vec{k} = \frac{\pi}{2a}(\hat{x}+\hat{y})$, $\vec{k} = \frac{3\pi}{a}\hat{x}$, and $\vec{k} = \frac{2\pi}{a}\left(\hat{x}+\frac{1}{2}\hat{y}\right)$, the curves split and form a photonic stop bands that indicates interactions between the surface plasmon polaritons and photonic modes hosted by the holes. The corresponding first Brillouin zone is shown in the inset of Fig. 2, where the three high symmetry points (M, $\Gamma$, X) are indicated. The decay rate $\Gamma$ of each mode is calculated as well and represented by color-coding the band diagrams. The decay rate corresponds to the optical lifetime $(\tau = 2\hbar/\Gamma)$ (Fig. 2b to d). At some specific energy ranges, one can clearly identify the high-quality factor PBM resonances,[41] in general though, a damping ratio of approximately 0.01 eV to 0.3 eV, corresponding to the life time in the range of 3.9 fs to 113.5 fs is recognized.

For hybrid $WSe_2$/Au-lattice structures, we consider a two-layer system for our simulations, composed of a $WSe_2$ thin film positioned on top of a plasmonic crystal (Fig. 2). It was recently reported[42] that for thin $WSe_2$ layers with thicknesses below 20 nm, due to the evanescent fields and the short propagation length of photonic modes in the $WSe_2$ flake, excitons do not interact strongly with photonic modes. When the interactions with plasmons are considered, in the case that the thickness of $WSe_2$ is 20 nm (Fig. 2b), the band structure inherits mostly the PBM properties and the plexcitonic group velocity ($v_g = \frac{1}{\hbar}\frac{dE}{dk}$) resembles that of the plasmonic crystal, although the PBMs at higher energies can interact with unscreened excitons (Fig. 2b). Upon introducing thicker $WSe_2$ films into our hybrid structures, we observe stronger exciton-plasmon interactions, which is partially due to the fact that the excitons that are excited at a certain distance from the metal are less screened by the free-electrons in the metal, and also the fact that the excitons themselves strongly interact with the photonic modes in thin $WSe_2$ films as well and form exciton polaritons.[43] Hence, despite the fact that the PBMs themselves sustain relatively short decay time, exciton polaritons in thicker flakes can propagate and interact with PBMs at a different lattice site. Therefore, thanks to the strong interaction of exciton polaritons with the PBMs, remarkably low group velocities of the emerging photonic bands are observed (Fig. 2c and d), that constitutes field profiles of increasing complexity by increasing the thickness of the $WSe_2$ flakes (compare field profiles A to H in Fig. 2). The group velocity of the Bloch modes in the hybrid crystal remains below 20% of the free-space light velocity, reaching zero values at certain points of the Brillouin zone (Supporting information, Fig. S6). For the wavelengths close to the excitonic absorption energies, electromagnetic energies are not transferred in the plexcitonic crystal (Supporting information, Fig. S1). Moreover, plexciton bandstructure and lifetime can be controlled precisely with the thickness of the $WSe_2$ flakes. Comparing Figures 2a and Fig. 2c at energies around 1.9 eV, the plexcitonic lifetime is larger than its quasi-propagating PBM counterparts. The numerical results mentioned above are used to interpret the experimental data that are described in the next section.

*Spectroscopic analysis of plexcitonic lattice* - The Au lattice, the $WSe_2$ flake, and the hybrid $WSe_2$/Au-lattice structures are all individually investigated experimentally, allowing for a detailed understanding of the responses of individual systems and their interactions. Cathodoluminescence spectroscopy is used to acquire hyperspectral energy-filtered images at the wavelength of 800 nm (Fig. 3a to c). We note that the energy-filtered images are a coherent superposition of all those optical modes with momenta positioned within the light cone that can couple to the electron beam. A swift electron propagating at the velocity $\vec{v}$ can interact only with those optical modes that sustain evanescent fields along the propagation direction of the electron, satisfying the momentum selection rule $\hbar \vec{k}_{\text{ph}} \cdot \vec{v} = \hbar \omega_{\text{ph}}$, where $\vec{k}_{\text{ph}}$ and $\omega_{\text{ph}}$ are the photon wave vector and angular frequency, respectively, and $\hbar$ is the reduced Planck's constant.[44,45] For our system, comprised of electrons at the kinetic energy of 30 keV, this corresponds to $k_{\text{ph}} = 3k_0$, that cannot be supported by the photonic modes confined to the holes.

Both for the plasmonic crystal structure and the hybrid WSe$_2$/Au-lattice structure, the hyperspectral images demonstrate a ring-like asymmetric maximum CL intensity surrounding the holes[45]. The asymmetric pattern is due to the reflection of the quasi-propagating PBMs from the boundaries of the truncated photonic crystal. CL spectra of the 70-nm-thick WSe$_2$ flake at selected electron impact positions demonstrate a wavelength split at the order of 160 nm centered at the A-exciton wavelength as the result of the exciton-photon interactions in the WSe$_2$ flake, which is in a good agreement with the recently reported values[28,46] and confirms exciton-polariton formation in WSe$_2$ thin films. Moreover, the hyperspectral image acquired at $\lambda = 650 \text{ nm}$ demonstrates the wave pattern of the exciton polaritons, revealed due to the interference of the scattered exciton polaritons from the edges of the flakes and the transition radiation (Fig 2a). The latter is excited when a moving electron traverses an interface, due to the interaction of the moving electron beam with its image charge inside the material. The CL spectra acquired at different electron impact positions for the Au lattice demonstrate a rather broad peak (Fig. 3e), centered at the wavelength of $\lambda = 800 \text{ nm}$, which is shifted ~60 nm with respect to the A exciton energy. The CL spectra sustain an inhomogeneous broadening, that is due to the excitation of a combination of plasmonic modes with different momenta and symmetries, as confirmed by the previously discussed simulation results.[41] The optical response of individual WSe$_2$ flakes and the formation of the exciton polaritons are thoroughly explored using CL spectroscopy elsewhere.[29] PBMs supported by the same plasmonic lattice have been also investigated with CL spectroscopy and angle-resolved mapping previously.[31,41,47] Here, we focus on the interaction of the PBMs with exciton polaritons.

The acquired CL spectra for the hybrid WSe$_2$/Au-lattice are distinct from the individual Au-lattice and WSe$_2$ flake spectra and show different behaviors (Fig. 2f): We observe three peaks when exciting the sample at the electron impact positions marked by A, B, and C, whereas for electrons impacting at the rim of the holes, only two peaks are observed. This behavior signifies the space-dependent coupling strength between excitons and plasmons, which happens due to the interaction of the excitons in WSe$_2$ with different PBMs. PBMs have the largest field enhancement factor at the rim of the holes (Fig. 3b and e, A site) and their interactions with A excitons, at this site, results in stronger exciton-photon interactions, therefore a wavelength splitting of 240 nm is observed, that is 80 nm more than the wavelength split observed for free-standing WSe$_2$ films (compare Fig. 3f D with Fig. 3d). The largest wavelength split is observed at site D (Fig. 3f D), corresponding with the interaction of the optical modes excited inside the holes (photonic cavity modes) with excitons. Although the PBMs hosted by the holes could not be directly resolved with electrons- due to the momentum selection rule described above, they strongly interact with excitons and the resulting energy split is revealed in the CL spectrum of the hybrid structure. Moreover, as stated above, three peaks at the wavelengths of 610 nm, 760 nm, and 890 nm are observed. This behavior is due to the strong interaction of PBMs with both A and B excitons. Indeed, as described above, the interaction of both A and B excitons result in the formation of energy band gaps in the band diagram of the hybrid exciton-plasmon lattice (Fig. 2), due to the hybridization of the PBM and exciton resonances, in such a way that bandgaps emerge around the wavelengths of $\lambda = 680$ nm and $\lambda = 780$ nm. This is the reason that at these wavelengths, a dip in the CL spectra is observed (green arrows in Fig. 3f).

To gain more insight into the coupling strength between excitons and PBMs, we use the effective Jaynes-Cumming Hamiltonian (see Supporting Information). We find a coupling strength of up to ~ 257 meV for A-excitons and ~ 70 meV for B-excitons corresponds to light matter interaction approaching the ultra-strong coupling regime for A exciton and a weak coupling with B exciton.

To investigate the role of the thickness of the WSe$_2$ layer on the emergence of the CL peaks originating from excited plexcitons in WSe$_2$/Au lattice hybrid structures, we look at the CL response of a double-layer WSe$_2$ system positioned on top of the plasmonic lattice. The WSe$_2$ in this case is made of two layers with 30 nm (flake 1) and 60 nm (flake 2) thicknesses respectively. The CL spectra are acquired at selected electron impact positions marked with yellow dots in the SEM image of Fig. 4(a).

CL spectra were acquired at different impact positions on this truncated hybrid structure (in the center of the hole, on the edge and in the space between two holes). When the electron beam interacts with the structure at regions outside the WSe$_2$ flakes, two broad peaks are observed. The appearance of these two peaks are understood by the excitation of the PBMs that interact with the edges of the WSe$_2$ flake, and that are partially transmitted into the WSe$_2$ film where they interact with the A exciton (see Supporting information, Fig. S3). The electron beam cannot excite the optical modes when they traverse the structure inside the holes at a distance larger than 10 nm away from the rim, as described above. When the electron beam excites the structure on flake 1, regardless of the electron-impact position, two peaks are observed. Hence, for flake 1, the exciton-plasmon interactions are not strong enough to form three bands (Fig. 4c), as confirmed by our numerical results (Fig. 2b, and for all thickness of the WSe2 flakes below 30 nm). Finally, when the electron beam traverses the structure through the thicker WSe$_2$ flake, again three hybridized bands are formed (Fig. 4(d)). The measured CL spectra agree well with the calculated CL spectra for which the exact topology of the system was considered. These data confirm the role of the thickness of the WSe$_2$ flake as a knob to control the strength of the plasmon-exciton interactions and to tune the dispersion of the emerging bands. By measuring the CL response of various flakes with different thicknesses, shifts in the energy of the CL peaks versus the thicknesses of flakes are observed (see supplementary Fig. S4 and S5). Comparing this to our numerical results outlined before, these behaviors are attributed to the emergence of flat bands and the dependence of their energies on the thickness of the WSe$_2$ flakes.

To investigate the band structure of the plexcitonic crystal in more detail, we performed angle-resolved energy-momentum CL measurements for the plasmonic crystals coated with a 50-nm-thick WSe$_2$ flake (Fig. 5). Energy-momentum CL maps are acquired along the direction marked by the red arrow shown in the insets of Fig. 5, and are acquired by filtering the radiated far-field excitations by a mechanical slit. These maps demonstrate an energy splitting, in agreement with the predicted bandgap energies of the numerically calculated band diagram of the same structure along the Γ→X and Γ→M directions in the reciprocal space (Fig. 5 a and c show the experimental results, and Fig. 5 b and d show the calculated band structure).

A pronounced gap near to the A-exciton energy (1.65 eV) is observed. This effect highlights strong plasmon-exciton interactions near to the energy of A-excitons and only weak interactions with the higher-energy B excitons can be observed. Exciton B has 3 times faster relaxation rates compared to exciton A[29]. This effect results in much shorter interaction times between Bloch plasmons and the B exciton. As a result, the resulting hybridized plexcitonic bands sustain larger damping ratio at higher energies, thus the broadening of the peaks observed in the experimental data does not allow for fully resolving the associated bandgap at higher energies. Thus, the dissipative losses are responsible for the broadening of the peaks observed in the experimental data and the linewidth of the optical bands in energy-momentum CL map.

These behaviors are fully reproduced in the numerically calculated band structure for a $WSe_2$ flake with 50 nm thickness (Fig. 5b and d). White dashed lines trace the dispersion of the waves caused by the scattering of the excited Cherenkov radiation from the edges of the $WSe_2$ flakes, which forms a strong signal in the acquired CL spectra (For more information see supplementary note 1). Since the Cherenkov radiation is quite broad band, it adds to the background of the acquired CL spectra integrated over the whole momentum space, hence does not lead to spectral peaks.

### 3. Discussion

In summary, through the extensive analysis of numerical and experimental measurements, we have shown that quasi-propagating plasmonic Bloch modes in plasmonic crystals strongly couple to excitons in semiconducting $WSe_2$ thin flakes. Herein, we have demonstrated the progressive enhancement in interactions from strong exciton-photon couplings in freestanding Wse$_2$ flakes up to ultra-strong coupling between the excitons of WSe$_2$ flakes and quasi-propagating plasmons supported by thin gold films including a square array of holes, forming quasi-propagating plexcitons with flat bands. Our comprehensive experimental measurements accompanied with numerical simulations, let us to reconstruct the full *k*-vector dependence of the dispersive properties of plexcitonic lattices. We also demonstrated that the dispersion of the plexcitons can be controlled by the thickness of the WSe$_2$ layer and the formation of flat-band polaritons. These results can directly impact many photonic systems and opens a pathway for realization of new kinds of topological photonic structures[48-49], by incorporating other forms of lattices that could support inverted bands. The remarkably low-group velocity of the flat-band optical waves could be used for enhanced light-matter interactions which is an interesting feature for nonlinear optics – though for these applications, incorporating strategies for reducing the attenuation constant – for example by using other forms of plasmonic lattices and materials - could be beneficial.

We indeed observe emission from exciton polaritons interacting with the edges of the flakes, not the excitons themselves, since excitonic luminescence yield in bulk Wse$_2$ is expected to vanish due to the indirect band gap of the multilayer structures compared to monolayers.[50] Hence, the broad CL spectra demonstrated here is understood by the dispersion of the exciton polaritons, as well as other mechanisms of radiation including transition radiation and the Cherenkov radiation, that latter has been also demonstrated to strongly interact with excitons.[51] Cherenkov radiation can only be excited in the bulk of the material and contributes to the differences observed in the CL spectra reported here compared to the CL investigations of ultrathin and monolayer flakes[52-53], next to the fact that exciton polaritons are not excited in the ultrathin flakes.

Furthermore, the plasmon-exciton coupling can control the effective mass of the emergent plexcitons.[54] The critical temperature of condensation and effective mass ($m_{eff}$) of the quasi particles, are key elements for hybrid TMDC/plasmonic crystal structures to reach the low threshold plexcitonic lasing[55] and observing condensation.[54] Moreover, by confining light in the form of flat-band plexcitons, one should be able to significantly change the photonic density of states of plexcitons ($n_c$) (see Supporting information, Fig. S6 and supplementary Note 2) which is a particularly fruitful platform for controlling the dynamics of the light-matter interactions as a crucial parameter for room temperature condensation (approximately $n_c \propto m_{eff}$)[56] and single-photon nonlinearities.[34] Aside from exploring TMDCs as a promising material for strong light matter coupling at room temperature, our hybrid plasmon-exciton structure can be easily integrated into optoelectronic devices such as solar cells[57] and novel compact photonic devices.[58,59]

All the TMDC flakes considered above are thick enough to neglect quantum corrections in the permittivity model arising from quantum confinements. These aspects, which normally happen in only atomically thin van der Waals materials have been not explored here, and could trigger further control knobs and emerging phenomena.

**Supporting Information**

The Supporting Information is available free of charge at xx.

**Acknowledgements**

The Authors acknowledge fruitful discussions with Dr. Andrea Konečná for simulations including electron beams. This project has received funding from the European Research Council (ERC) under the European Union's Horizon 2020 research and innovation programme, Grant Agreements No. 802130 (Kiel, NanoBeam) and Grant Agreements No. 101017720 (EBEAM). Financial support from Deutsche Forschungsgemeinschaft under the Art. 91 b GG Grant Agreement No. 447330010 and Grant Agreement No. 440395346 is acknowledged.

**Data availability**

The data that support the findings of this study are available from the corresponding author upon reasonable request.

**Conflicts of interest**

Toon Coenen is an employee of Delmic B.V., a company that develops and sells the cathodoluminescence system that was used in this paper.

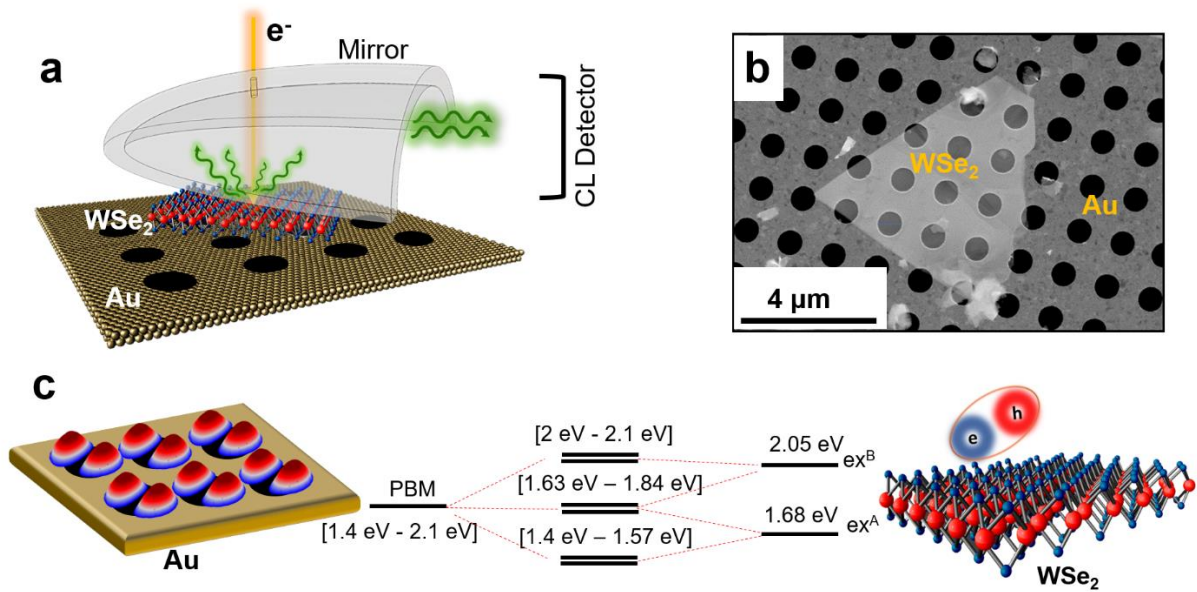

**Figure 1: Strong exciton-plasmon interactions in a hybrid plexcitonic lattice.** (a) Schematic overview of the experimental setup inside a scanning electron microscope (SEM). The plexcitons emerging from the interaction of excitons in the WSe$_2$ flake with Bloch plasmons in the gold lattice are excited by an electron beam and probed using cathodoluminescence spectroscopy. (b) High-magnification SEM image of the WSe$_2$ thin flake deposited on top of the two-dimensional square plasmonic crystal with periodically patterned hole arrays in a 50 nm-thick Au slab. The square lattice constant is 1.6 µm and the hole diameter is 0.8 µm. (c) Schematic illustration of emerging hybridized bands and plexciton quasiparticle creation via the strong coupling between plasmonic Bloch modes (PBMs) with A- and B-excitons of the WSe$_2$ flake. Demonstrated on the left is the spatial profile of a specific mode of the plasmonic lattice; namely, the dipolar excitations. The interaction between excitons and plasmonic Bloch modes result in three typical bands, for a given flake thickness of 60 nm.

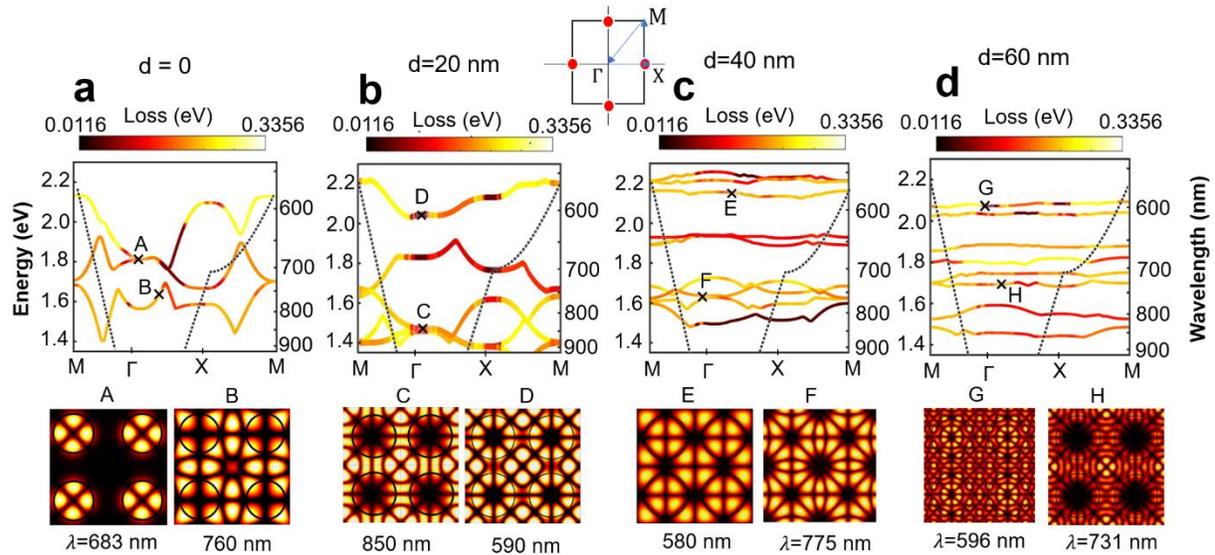

**Figure 2: Calculated band structure of the plexcitonic lattice.** (a) The band diagram of the gold square lattice plasmonic crystal, without a Wse₂ layer on top. (b)-(d) Calculated band structure of the hybrid plasmonic crystal/WSe₂ flake for WSe₂ thicknesses of 20 nm, 40 nm, and 60 nm, respectively. The inset shows the reciprocal space lattice sites of the plasmonic crystal with the square lattice (red circles) and the blue arrows show the direction in the reciprocal space (M-Γ-X-M) adopted for calculating the dispersion diagram. The dashed line represents the optical line associated with plane waves in a square lattice. The decay rate of the plasmonic and the plexcitonic hybrid modes are presented by color coding the dispersion diagrams. Spatial profiles of specific modes are shown in the panels below the band diagrams. For better comparison, the spatial profiles of the resonance modes are selected near Γ point in different bands. The depicted modes are also marked as crosses in panels (a) to (d).

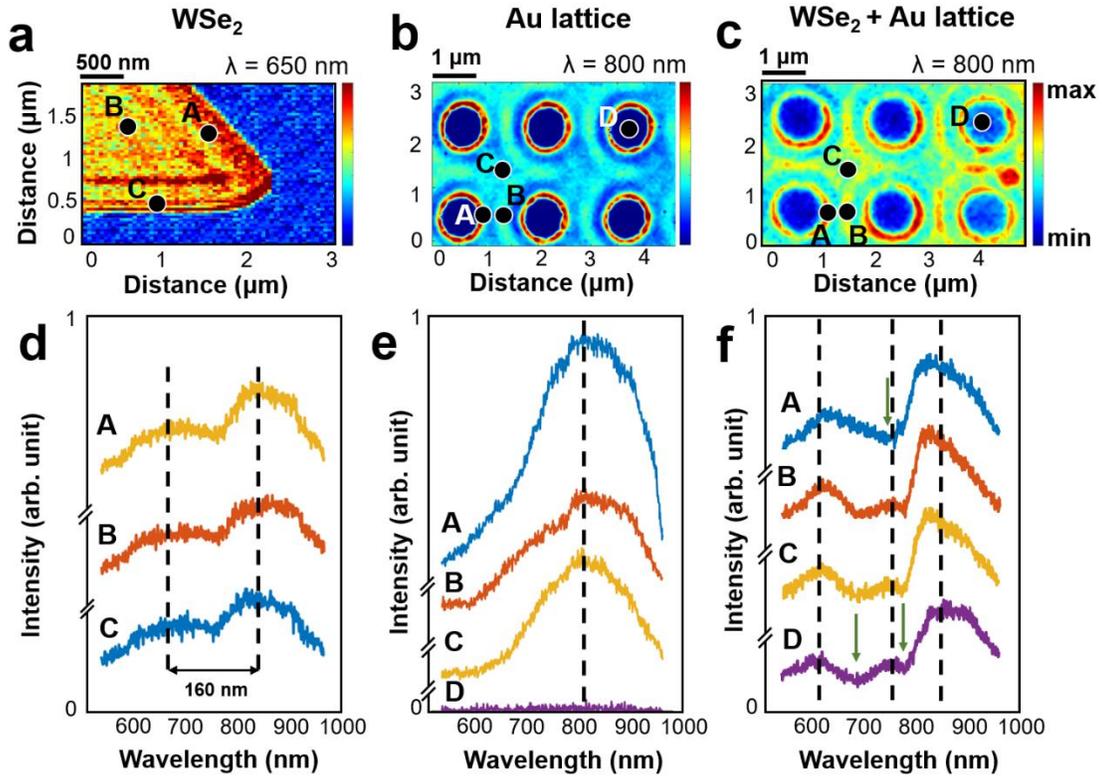

**Figure 3: Cathodoluminescence spectroscopy of the plexcitonic crystal**. (a) The spatial distribution of the cathodoluminescence signal in (a) a triangular WSe$_2$ flake with the thickness of 70 nm at a wavelength of λ=650 nm averaged over 20 nm bandwidth, where the interference fringes caused by the reflection of exciton polaritons from the edges of the triangle is apparent, (b) the plasmonic lattice, and (c) the hybrid plexcitonic lattice at the wavelength of λ=800 nm averaged over 20 nm bandwidth. (d) Measured CL spectra at selected electron impact positions as indicated in panel (a). The strong interaction between excitons and photons in the WSe$_2$ thin film causes a 160-nm wavelength split at the A-exciton energy, as is clear from the measured CL spectra. Measured CL spectra at selected electron excitation positions in (e) gold lattice and (f) WSe$_2$/plasmonic hybrid crystal, at electron impact positions marked in panels (b) and (c), respectively. The green arrows indicate dips that demonstrate the strong plasmon-exciton interactions in the plexcitonic lattice, since they highlight the formation of the energy gaps in the band diagram shown in Fig. 2. The strong deviation of the spectral shapes of the plexcitonic lattice compared to the Wse$_2$ flakes and the plasmonic lattice, demonstrate the significant role of quasiparticle interactions in forming the optical response of the whole structure.

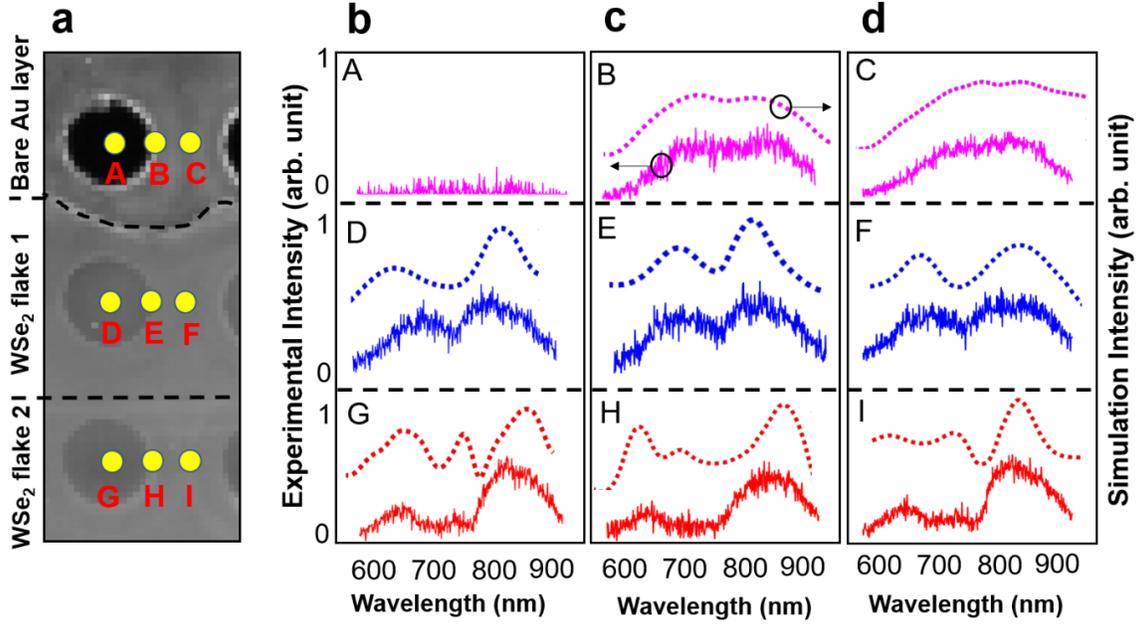

**Figure 4: The role of the thickness of the Wse₂ film on the emerging CL peaks.** (a) The SEM image of the WSe₂ structure positioned on top of the plasmonic lattice. The WSe₂ structure is made of two layers with 30 nm (flake 1) and 60 nm (flake 2). The yellow dots depict the selected electron excitation impact positions used to acquire the CL spectra. (b)-(d) Measured (solid lines) and Simulated (dotted lines) CL spectra at the selected electron impact positions, as indicated in panel (a) in the hole center, on the edge and in the space between two holes. The CL spectra shown in sub-panels A-C, are originated from quasi-guided plasmonic Bloch modes excited by the plasmonic crystal. Strong interactions between plasmons and both exciton A and B excitations can occur only in thicker flakes, where the excitons are not fully screened by the conduction electrons in metals.

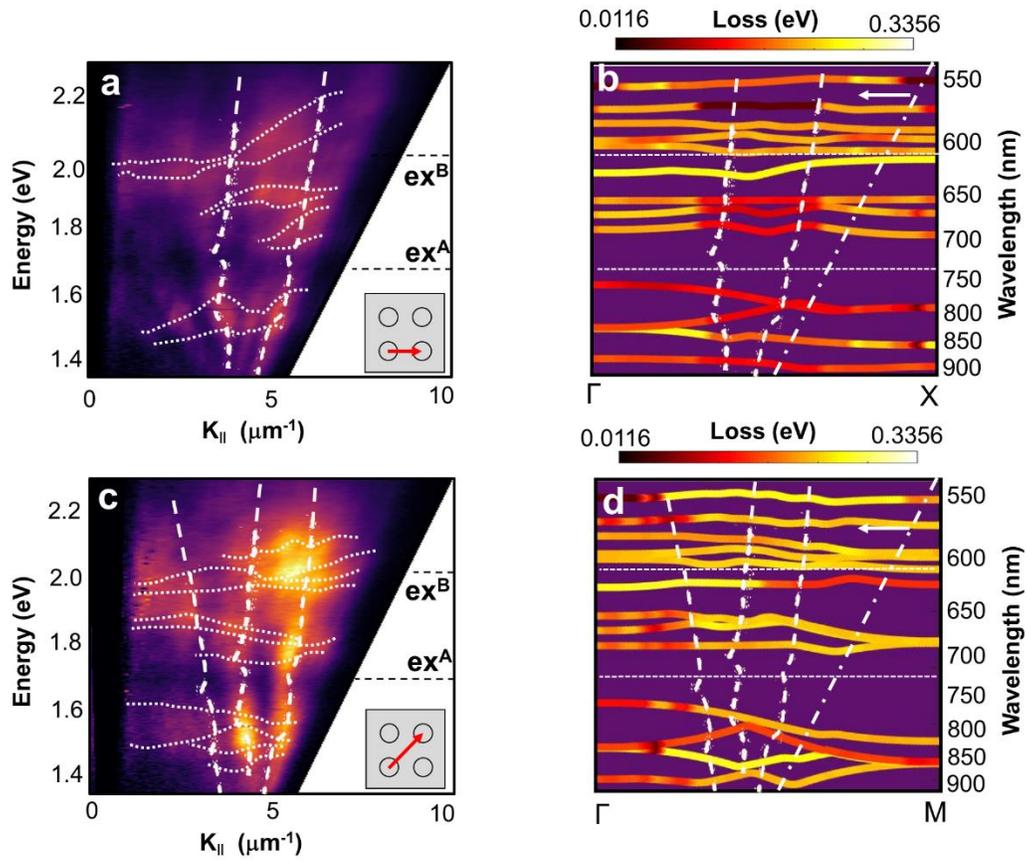

**Figure 5: Probing the band structure of the plexcitons in hybrid Au/WSe₂ crystal.** (a, c) Measured energy-momentum CL map, where the momentum direction along the red arrow shown in the inset schematics are filtered by a mechanical slit positioned in the far-field, respectively. White dotted lines traces the observed momentum resolved peaks and were used to highlight the formation the plexcitonic branches. Calculated band structures along the (b) Γ→X and (d) Γ→M directions in the reciprocal space for the hybrid Au/WSe₂ structure, where the Wse₂ flake has a thickness of 50 nm. The oblique dashed-dotted lines represent the optical lines associated with plane waves in a square lattice. Only the regions within the light cone could be experimentally unraveled. White dashed lines in panels trace the dispersion of the waves caused by the interference of scattering from the edges of the WSe₂ flakes with the transition radiation at the far field. Exciton energies are indicated by dotted lines.